\documentclass[a4paper]{article}

\pdfoutput=1

\usepackage{a4wide}
\usepackage{graphicx}
\usepackage{lscape}
\usepackage{hyperref}

\begin{document}

\title{Fast Determination of Constellation Membership}
\author{Patrick Glaschke}
\maketitle

\begin{abstract}
The 88 constellations as defined by the IAU segment the sky into regions, separated by
an intricate set of boundaries. A binary tree decomposition of this landscape is given 
which tessellates the celestial sphere into rectangles. This allows a fast determination of
the constellation membership of any given sky coordinate.

\vspace*{3mm}

\end{abstract}

Determining the constellation membership of a celestial coordinate is a tedious task: Though
the constellation boundaries follow lines of constant declination and right ascension at epoch
B1850.0, they take a winding course over the sky to assure consistency with older membership
definitions of various sky objects.

A first step beyond the brute-force check of every constellation is a search in a boundary set
ordered by declination and right ascension. Roman \cite{roman1987identification} has prepared such 
a data table which allows the user to stop the identification process as soon as a positive 
detection has occurred without
processing the remaining constellations. However, the worst case still requires an
exhaustive search with a total amount of more than 1,000 coordinate comparisons. This might
be prohibitive if a mass classification of celestial objects or an interactive classification 
in real time is desired.

Several options are available to speed up the search. The most direct approach is to 
tessellate the sky by a rectilinear mesh such that every mesh cell has a unique constellation 
membership. While the query of a given cell is $\mathcal{O}(1)$, determining the mesh cell from a celestial
coordinate introduces some overhead. Furthermore, some 40,000 data entries need to be
stored which is far more than the original boundary data.

A more indirect approach followed in this paper is the partitioning of the sky into a binary tree. 
Trees are known for the efficient structuring of data, though the complicated layout of the
constellation boundaries asks for special adaptations. The construction of a binary partition
is not unique and allows for the optimization of different goals. A hierarchical division into
portions of equal areas minimizes the average search depth, while the worst case performance
is not controlled. Splitting the data into sets with an equal number of boundary segments yields a better
worst case performance, but is on average inferior to the afore mentioned solution.
As the aim of this work is a fast and reliable solution for any location on the sky, 
the second approach is preferred.

Having fixed the general construction rule of the binary tree, the remaining task is to
conduct the subsequent partitioning of the boundary set. The initial boundary data is
taken from Davenhall \cite{Davenhall1989} except the additional boundaries for Octans introduced
by him. All adjacent collinear boundary edges are merged and edges crossing the
$0^{\mathrm{h}}$ meridian are split to assure the reliable detection of candidate splits.

\newpage 

The algorithm used works as follows:
\begin{enumerate}
\item
  Every boundary segment in the current data set is extended to form a \textit{candidate split}.
\item
  The current data is divided according to each candidate split while the number of segments in 
  each partition is recorded.
\item \label{BalanceCrit}
  The split which minimizes the size of the largest partition is chosen as the final split line. 
  This selection avoids strongly imbalanced trees while avoiding unfavorable cuts at the same time.
\item
  The boundary segments are divided according to the selected split. Segments on the split line are discarded,
  segments intersecting the split line are separated into two segments.
\item
  Each data set is processed recursively until empty. The constellation membership of the corresponding 
  partition is derived from the last split line.        
\end{enumerate}
It must be noted that this heuristic can not replace a rigorous minimization of 
the maximal tree depth, but merely aims at avoiding worst case scenarios. The resulting tree allows
the constellation classification with at most 11 decisions, though most queries are resolved with
9 or 10 decisions.

\begin{landscape}
\begin{figure}
\centering
\includegraphics*[scale=0.75,angle=0]{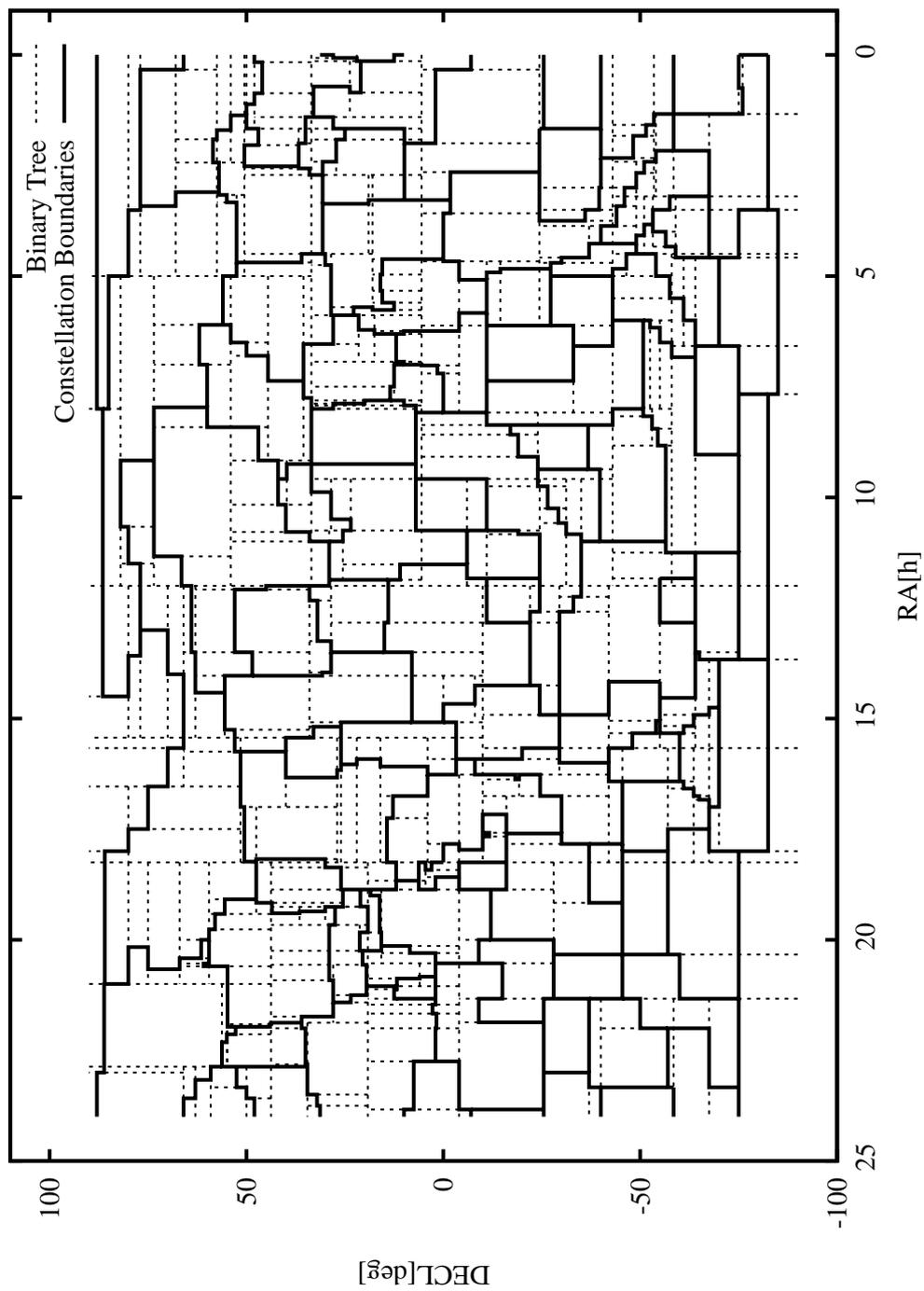}
 \caption{Tesselation of the constellation boundaries. \label{TreeBounds}}
\end{figure}
\end{landscape}

Fig. \ref{TreeBounds} visualizes the constellation boundaries in B1875.0 coordinates as well as the resulting
binary tree. Note how the binary cuts are alinged with the constellation boundaries. Especially Ursa Minor and 
Octans are victims of the tree balancing criterion. These constellations could be
tessellated with a rather small number of cuts, but such splits are discarded to maintain an overall balanced tree.
A self-contained C implementation of the constellation search is given as an ancillary file.

\bibliography{Constellation}
\bibliographystyle{plain}
\nocite{*}                      

\end{document}